\documentclass[12pt]{article}
\usepackage{amssymb,amsmath,epsfig}
\begin{document}
\parindent=0pt
\parskip=6pt
\rm
\begin{center}
\textbf{Application of two-sublattice bilinearly coupled Heisenberg
	model to the description of certain ferrimagnetic materials}\\
\end{center}
Hassan Chamati
chamati@issp.bas.bg\\
and Diana V. Shopova
sho@issp.bas.bg
Corresponding author\\
Address:Institute of Solid State Physics, Bulgarian
	Academy of Sciences, 1784 Sofia, Bulgaria

Keywords;
ferrimagnetism, Landau theory, Heisenberg model, phase diagram.

\begin{abstract}
We study phenomenologically on the basis of two bilinearly coupled
Heisenberg models the phase diagram of some ferrimagnetic
substances. Calculations are performed with the help of Landau
energy obtained through applying the  Hubbard-Stratonovich
transformation to the initial microscopic Heisenberg
Hamiltonian. The phase transitions within the model  are of
second order with the emergence of a compensation point at lower
temperatures for some values of parameters of the system. The main phase
is a two-sublattice collinear ferrimagnet but also a
metastable non-collinear phase is  present within the exchange
approximation  presented here. The numerical results give a
detailed description of temperature dependence of
magnetization on the strength of intersublattice interaction
and the difference between the effective exchanges of two
ferromagnetically ordered sublattices.
\end{abstract}

\section{\label{1} Introduction}
Ferrimagnets are substances made of various components having
different magnetic properties. The differences in magnetic moments lead
to a geometric frustration that may
arise because either different elements occupy the lattice sites or
the same element occupies nonequivalent crystallographic sites
surrounded by a different number or type of non-magnetic ions, which
effectively results in different magnetic properties. For complex
alloys a combination of both may take place (For an extensive review
see Ref. \cite{skomski:2008} and references therein). Within mean-field approach it is generally
accepted that ferrimagnets can be modeled with the help of several
interpenetrating sublattices each ordered ferromagnetically with
effective antiferromagnetic coupling between them. In the pioneering
works of N\'{e}el on ferrimagnetism within the molecular field
approach (see e.g. \cite{Neel:1971}), a two sublattice model is used to compute
the thermal magnetization behaviour of ferrimagnets, and six possible
magnetization curves were derived. Special attention there is paid to
iron garnets, where the spontaneous magnetization in comparison with
experiment can be interpreted by applying a three-sublattice model. In
view of experimental study of magnetocaloric effect of rare-earth
based ferrimagnets~\cite{Andreenko:1989} which has great potential
for technological applications in environmentally-friendly
refrigeration, the theoretical mean-field description of such alloys
with three sublattices, is further elaborated~\cite{Ranke:2009}. Such
studies are based on considering microscopic classical Heisenberg
models with different exchange and spin-orbit interactions depending
on the crystal structure, chemical composition of the particular alloy
under study. \\There is another theoretical mean field approach based on
considering mixed spin Ising model for description of ferrimagnets,
see for example~\cite{Kaneyoshi:1991,Tucker:1999,Abubrig:2013}, where a very
detailed review of the literature on this approach is presented.
In the present paper we will consider ferrimagnets which can be described
by different magnetic ions sitting on two interpenetrating
sublattices in a body centred cubic structure. The interaction
of ions on each sublattice is supposed ferromagnetic, while ions on the different sublattices are coupled
antiferromagnetically. The magnetic properties will be investigated on
the basis of bilinearly coupled Heisenberg classical model in a mean-field
approximation which is treated using the Hubbard-Stratonovich
transformation for obtaining the respective Landau free energy and its
analysis.

The rest of the paper is organized as follows: in Section 2 we describe in detail how we calculate the Landau free energy from classical Heisenberg model with competing interactions on the basis of previously applied approach~\cite{Shopova:2001} for derivation of mean field approximation. In Section 3 the solutions of equations of state obtained by the minimization of Landau energy derived in Section 1 are discussed. Section 4 summarizes the results both in strong and weak-coupling limit for ferrimagnetic substances under consideration.  Section 5 generalizes the conclusions and possible further development of our study.

\section{\label{2} The model and derivation of Landau free energy}
 The microscopic Heisenberg Hamiltonian which describes two coupled subsystems consisting of classical spins with different magnetic properties which antiferromagnetically between them through a bilinear term can be written in the following form :
\begin{equation}\label{Eq1}
H=-\frac{1}{2}\sum_{ij}^{2N} \left[\mathcal{J}^{(1)}_{ij}\mathbf{S}_i^{(1)}
\cdot \mathbf{S}_j^{(1)} + \mathcal{J}^{(2)}_{ij}\mathbf{S}_i^{(2)} \cdot
\mathbf{S}_j^{(2)} +2 \mathcal{K}_{ij}\mathbf{S}_i^{(1)} \cdot
\mathbf{S}_j^{(2)}\right].
\end{equation}
Here $\mathbf{S}_j^{(1,2)}$, are $n$-component classical Heisenberg spins  whose magnitude is normalized on the
unit sphere in spin space through the condition $|\mathbf{S}_j^{(1,2)}|=1$. The exchange parameters $\mathcal{J}^{(1,2)}_{ij}, \; \mathcal{K}_{ij}$ in the general case are $N \times N$ symmetric matrices with $N$ - the number of lattice sites considered equal for both subsystems. This condition simplifies the consideration as makes the system symmetric with respect to the interchange of the subsystems. The exchange matrices $ \mathcal{J}^{(1,2)}_{ij}$ denote the interaction between magnetic atoms of the same sort and  $\mathcal{K}_{ij}$ - between magnetic atoms of different sorts. \\The above Hamiltonian may be applied to the description of magnetic systems which consist of two different magnetic materials and no matter what is the microscopic origin of this  difference, it is effectively described  by different exchange interactions within the two subsystems. There may be other situation when the substance is made only of one type of  magnetic ions but they occupy two different crystallographic positions in the Bravais lattice and are separated by a number of non-magnetic atoms. Such substance may also be considered as built of two magnetic subsystems with different exchange interactions within them.\\
In order to analyse the behaviour of magnetization and the phase transitions in systems that can be described by the above microscopic Hamiltonian we have to find the mean-field energy for the Hamiltonian~(\ref{Eq1}) by calculating the partition function which in this case is represented by functional integral in n-dimensional spin space, where $n$ - is the number of spin components.  To do this we apply the Hubabrd -Stratonovich transformation; see for example~\cite{Brout:1974} and the papers cited therein. We have used this approach in \cite{Shopova:2001} for ferromagnetic coupling between the two magnetic subsystems where a detailed description of procedure is given. Here we will just outline the important steps in the derivation of Landau free energy , especially in relation of antiferromagnetic coupling between the subsystems. \\
We present the two interacting different magnetic subsystems on a body-centered crystal lattice, for which the corners of elementary cube are occupied by one sort of magnetic atoms,  and at the center of the cube atoms of different magnetic sort are located. So the nearest neighbours belong to different magnetic subsystems and the next-nearest neighbour to subsystems 1 and 2,  respectively. Thus the system may be described  as two interpenetrating sublattices,consisting of different magnetic atoms and we assume that the interaction within the sublattices $\mathcal{J}^{1,2}_{ij}$ is ferromagnetic and between them, $\mathcal{K}_{ij}$,  it is antiferromagnetic.
The Hubbard-Stratonovich transformation renders the initial microscopic Hamiltonian in new n-component variables $\mathbf{\Psi}^{(1)}_i, \mathbf{\Psi}^{(2)}_i$  defined in real space, directly connected with the initial spins (see~\cite{Shopova:2001}), namely:
\begin{align}\label{Eq2}
\mathcal{H}=\frac{1}{2}\sum_{ij}^{2N}\left(J_{ij}^{(1)}\mathbf{\Psi}_i^{(1)} \cdot \mathbf{\Psi}_j^{(1)}+ J_{ij}^{(2)}\mathbf{\Psi}_i^{(2)} \cdot \mathbf{\Psi}_j^{(2)}+2K_{ij}\mathbf{\Psi}_i^{(1)} \cdot \mathbf{\Psi}_j^{(2)}\right) \\
\nonumber- \ln\left[\sum_{i}^{2N}\textsl{I}_{n/2-1}(x_i^{(1)}) (\frac{x_i^{(1)})}{2})^{-n/2}\Gamma(\frac{n}{2})\right] - \ln\left[\sum_{i}^{2N}\textsl{I}_{n/2-1}(x_i^{(2)} )(\frac{x_i^2}{2})^{-n/2}\Gamma(\frac{n}{2})\right].
\end{align}
Here $\textsl{I}_{n/2-1}(x_i^{(1,2)} )$ is the modified Bessel function, and $\Gamma(\frac{n}{2})$ is the Gamma function. In the above expression the exchange parameters $J_{ij}^{(1,2)}$ and $K_{ij}$ are connected to those in the initial Hamiltonian ~(\ref{Eq1}) by the relations:
\begin{eqnarray}\label{Eq3}
\nonumber  J_{ij}^{(1,2)} &=& \frac{\mathcal{J}^{(1,2)}_{ij}}{T} \\
  K_{ij} &=& \frac{\mathcal{K}_{ij}}{T}
\end{eqnarray}
with $T$ - the temperature. \\ We denote by  $x_i^{(1)}$ and $x_i^{(2)}$ in (\ref{Eq2}) the following expressions:
$$x_i^{(1)}=\left|J_{ij}^{(1)}\mathbf{\Psi}_j^{(1)}+K_{ij}\mathbf{\Psi}_j^{(2)}\right|; \;\; x_i^{(2)}= \left|J_{ij}^{(2)}\mathbf{\Psi}_j^{(2)}+K_{ij}\mathbf{\Psi}_j^{(1)}\right|$$
The terms containing Bessel functions in~(\ref{Eq2}) will be further used only in the form of expansion with respect to $x_i^{(1,2)}$ up to forth order by using the relation:
$$\Gamma(\frac{n}{2}) (\frac{2}{x})^{(n/2-1)}I_{n/2-1}(x)=1+\sum_{k=1}^{\infty}\frac{(x^2/4)^k}{k!(k+n/2+1)(k+n/2-2) ...n/2}$$
The next step is to perform  Fourier transformation to $\mathbf{k}$-space , and pass to continuum limit in $\mathbf{k}$ as the finite size effects will not be considered at this stage.
The quadratic part of the obtained Hamiltonian again contains a bilinear term with respect to $\mathbf{\Psi}^{(1,2)}(\textbf{k})$ and we have to diagonalise it. This is done with the  help of unitary matrix $\hat{S}$:
\begin{equation}\label{Eq4}
\hat{S}= \left(
\begin{array}{cc}
S_0(\mathbf{k}) & -S_1(\mathbf{k}) \\
S_1^{\ast}(\mathbf{k}) & S_0(\mathbf{k}) \\
\end{array}
\right).
\end{equation}
The eigenvalues of the matrix $\hat{S}$ read:
\begin{equation}\label{Eq5}
\lambda_{1,2}(\mathbf{k})=\frac{1}{2}\left[J_1(\mathbf{k})+J_2(\mathbf{k}) \pm \sqrt{(J_1(\mathbf{k})-J_2(\mathbf{k}))^2+4K(\mathbf{k})^2}\right],
\end{equation}
where $J_{1,2}(\mathbf{k})$ and $K(\mathbf{k})$
are the Fourier transforms of $J_{ij}^{(1,2)}$ and
$K_{ij}$, respectively. In order to compute the
integral  we use the steepest-descent method, \textit{i.e.} the
integration contour is taken around the maxima of the
eigenfunctions (\ref{Eq5}).  The calculation for bcc structure show that if we take the nearest neigbour interaction between atoms of the same sort  and the nearest neighbour interaction between the atoms of different sort, $\lambda_{1,2}(\mathbf{k})$ has a maximum in the centre of the
Brillouin zone that gives ferromagnetic  ordering for the sublattices with antiferromagnetic $K<0$ interaction between them
which is the focus of our consideration below.  We
should note that $\lambda_{1,2}(\mathbf{k})$  has a maximum also at
the border of the Brillouin zone $\mathbf{k}=\pi/a$ ($a$ is the
lattice constant) which supposes antiferromagnetically ordered
sublattices. There may be also some local maximum inside the Brillouin
zone, which may give some incommensurate ordering within the sublattices,
but this case is beyond the scope of the present study.\\
After performing the reverse Fourier transform to real space we obtain the
dimensionless Landau energy in the following form:
\begin{align}\label{Eq6}
	\frac{F}{T} = \frac{t_1}{2}\overrightarrow{\psi}_1^2 & +\frac{t_2}{2}\overrightarrow{\psi}_2^2+\frac{g}{4}\left[(\overrightarrow{\psi}_1^2)^2+(\overrightarrow{\psi}_2^2)^2\right]
 \nonumber\\
& +\frac{b}{2}\overrightarrow{\psi}_1^2\overrightarrow{\psi}_2^2
+b(\overrightarrow{\psi}_1\cdot\overrightarrow{\psi}_2)^2+w(\overrightarrow{\psi}_2^2-\overrightarrow{\psi}_1^2)(\overrightarrow{\psi}_1\cdot\overrightarrow{\psi}_2).
\end{align}
The coefficients of the Landau energy are expressed by  the components of the matrix~(\ref{Eq4}) and its eigenvalues~(\ref{Eq5}) for $\mathbf{k}=0$:
\begin{align}\label{Eq7}
\nonumber S_0 & =  \frac{1}{D}\left(J_1-J_2 +\sqrt{(J_1-J_2)^2+4K^2}\right), \\
S_1 & = \frac{2K}{D}
\end{align}
where $D$ is introduced to satisfy the condition $\|\hat{S}\|=1$,
namely $S_0^2+S_1^2=1$:
\begin{equation}\label{Eq8}
D=\sqrt{2}\left[(J_1-J_2)^2+4K^2\right]^{1/4}\left[J_1-J_2 +\sqrt{(J_1-J_2)^2+4K^2} \right]^{1/2}.
\end{equation}
We will write here the explicit expressions for the coefficients of landau energy as we will need them further in solving the mean field equations and discussion of obtained results;
\begin{subequations}\label{Eq9}
\begin{align}
t_{1,2} & = \frac{1}{\lambda_{1,2}}-\frac{1}{n}\\
g & = \frac{u}{2}(S_0^4+S_1^4) \\
b & = uS_1^2S_0^2 \\
w & = \frac{u}{2}S_0S_1(S_0^2-S_1^2);
\end{align}
\end{subequations}
here $u=n^2 (n+2)$ with $n$  - the number of order parameter components.
The real vector fields $\overrightarrow{\psi_1}$ and $\overrightarrow{\psi_2}$ in the Landau free energy~(\ref{Eq6}) play the role of two coupled order parameters, and the averaged sublattice magnetizations are related to them by the equations:
\begin{subequations}\label{Eq10}
\begin{align}
\overrightarrow{m_1} & = \frac{S_0}{\lambda_1}\overrightarrow{\psi_1} -\frac{S_1}{\lambda_2}\overrightarrow{\psi_2}\\
\overrightarrow{m_2} & = \frac{S_1}{\lambda_1}\overrightarrow{\psi_1} +
\frac{S_0}{\lambda_2}\overrightarrow{\psi_2}
\end{align}
\end{subequations}

\section{\label{3} Solving mean-field equations}

The initial microscopic Hamiltonian is symmetric with respect to the rotation of all spins through the same angle. The application of Hubbard-Stratonovich transformation for derivation of landau free energy , given in previous section, preserves the  symmetry of initial hamiltonian also with respect to field variables $\overrightarrow{\psi}_{1,2}$ which means that we can find the magnitude and the mutual orientation between order parameters $\overrightarrow{\psi}_{1,2}$ but not their orientation with respect to crystallographic axes. This may be done for particular magnetic substance by including in the initial microscopic Hamiltonian terms accounting for the magnetic anisotropy. For pure exchange interactions we can introduce the following notations~\cite{Shopova:2001}:
 $\overrightarrow{\psi_1}_i=|\overrightarrow{\psi_1}|\beta_i$ and
 $\overrightarrow{\psi_2}_i=|\overrightarrow{\psi_2}|\delta_i$, where
 $|\overrightarrow{\psi_1}|=\psi_1, \;
 |\overrightarrow{\psi_2}|=\psi_2$ are the magnitudes of the vector
 fields, and $\beta_i, \; \delta_i$ are the respective direction
 cosines, which fulfil the condition:
\begin{equation}\label{Eq11}
\sum_{i=1}^3\beta_i^2=1 \quad \mathrm{and} \quad
	\sum_{i=1}^3\delta_i^2=1.
\end{equation}
The equations of state then will be:
\begin{equation}\label{Eq12}
 \frac{\partial f}{\partial X_i}=0,
\quad
\mathrm{where}
\quad
X_i=\{\psi_1, \psi_2, \beta_i, \delta_i\}
\end{equation}
Solving the above equations with respect to direction cosines $\beta_i, \; \delta_i$ gives two possible orientations between the vector fields $\overrightarrow{\psi_1}, \overrightarrow{\psi_2}$ for $K<0$:
\begin{enumerate}
\item The collinear phase with $\sum_i \beta_i \delta_i = -1$,
that is, $\overrightarrow{\psi_1}$ and $\overrightarrow{\psi_2}$ are antiparallel, and
\item The non-collinear phase with $\sum_i \beta_i \delta_i =
0$, that is, $\overrightarrow{\psi_1}$ and $\overrightarrow{\psi_2}$ are perpendicular.
\end{enumerate}
Below we will discuss in detail the non-collinear phase 2. The angle
between the order parameters $\overrightarrow{\psi_1}$ and
$\overrightarrow{\psi_2}$, \emph{i.e.},is;
$$
\sum_i^3 \beta_1 \delta_i = -\frac{w (\psi_2^2-\psi_1^2)}{2 b \psi_1 \psi_2}
$$
and is defined only when $\psi_1\neq 0$ and  $\psi_2\neq 0$. For $K <
0$, the analysis shows that the non-collinear phase exists only when the order parameters $\overrightarrow{\psi_1}$ and
$\overrightarrow{\psi_2}$ are of equal magnitudes, meaning that the
order parameters $\overrightarrow{\psi}_1 $ and
$\overrightarrow{\psi}_2$ are mutually perpendicular. The magnitude
$\psi=\psi_1=\psi_2$ for the non-collinear phase in analytical form
reads:
\begin{equation}\label{Eq13}
\psi^2=-\frac{t_1+t_2}{u}.
\end{equation}
Then the sublattice magnetization magnitudes calculated
using the above expressions for the non-collinear phase will be:
\begin{subequations}\label{Eq14}
\begin{align}
|\overrightarrow{m}_1| & =
	\psi\sqrt{\frac{S_0^2}{\lambda_1^2}+\frac{S_1^2}{\lambda_2^2}}, \\
|\overrightarrow{m}_2| & =
	\psi\sqrt{\frac{S_1^2}{\lambda_1^2}+\frac{S_0^2}{\lambda_2^2}}.
\end{align}
\end{subequations}
Note that the sublattice magnetizations are not perpendicular but
form an angle $\angle(\overrightarrow{m}_1,\overrightarrow{m}_2)
=\gamma$ with each other, expressed by
$$
\cos(\gamma)=\frac{S_0S_1(\lambda_1^2-\lambda_2^2)}{\sqrt{(S_0^2\lambda_2^2+S_1^2\lambda_1^2)(S_1^2\lambda_2^2+S_0^2\lambda_1^2)}}.
$$
The calculations show that this non-collinear phase for $K<0$ within the exchange approximation has no domain of stability. We should mention here that the free energy ~(\ref{Eq6}) is very sensitive to the sign of interaction $K$ between the sublattices. When $K>0$, \emph{i.e.}, the interaction between the sublattices is ferromagnetic there is small domain in which the respective non-collinear phase is stable \cite{Shopova:2003}.\\
For antiparallel $\overrightarrow{\psi_1}$ and
$\overrightarrow{\psi_2}$ it is obvious that the sublattice
magnetizations~(\ref{Eq10}) will be also antiparallel. We may write the
resulting equations for the magnitudes of the order parameters
$\psi_1$ and $\psi_2$ of the collinear phase and $K<0$ in the
following form:
\begin{subequations}\label{Eq15}
\begin{align}
t_1\psi_1 + g \psi_1^3 + 3b \psi_1\psi_2^2-w \psi_2
	(\psi_2^2-3\psi_1^2) & = 0, \\
t_2\psi_2+ g \psi_2^3 + 3b \psi_1^2\psi_2-w \psi_1
	(3\psi_2^2-\psi_1^2) & = 0,
\end{align}
\end{subequations}
with the stability conditions given by:
\begin{align}\label{Eq16}
	t_1+3g\psi_1^2+3b\psi_2^2+3w\psi_1\psi_2 & > 0 \\
	(t_1+3g\psi_1^2+3b\psi_2^2+6w\psi_1\psi_2)&(t_2+3g\psi_2^2+3b\psi_1^2-6w\psi_1\psi_2)\nonumber\\
	-9[w(\psi_1^2-\psi_2^2)+2b\psi_1\psi_2]^2 & \geq 0
\end{align}

We will make some remarks on the dependence of solutions of above system on the magnitude of exchange parameters $J_1, \; J_2$ and $K$. When $J_1<|K|, J_2<|K|$, the leading interaction is determined by the antiferromagnetic coupling between the two sublattices. This may be called a strong coupling limit for which the eigenvalue $\lambda_2(\mathbf{k}=0)$
\begin{equation}\label{Eq16}
\lambda_{2}=\frac{1}{2}\left[J_1+J_2-\sqrt{(J_1-J_2)^2+4K^2}\right],
\end{equation}
becomes negative. This is equivalent to the inequality $K^2-J_1J_2 >0$. The coefficient $t_2$ in front of $\psi_2^2$ becomes  positive; see (\ref{Eq9}), and the field $\overrightarrow{\psi}_2$  becomes redundant. The Landau free energy (\ref{Eq6}) will be:
\begin{equation}\label{Eq17}
 (\frac{F}{T})_s= f_s = \frac{t_1}{2}\overrightarrow{\psi_1}^2 +\frac{g}{4}(\overrightarrow{\psi_1}^2)^2
\end{equation}
The minimization of above equation gives for $\overrightarrow{\psi_1}$ the solution:
\begin{equation}\label{Eq18}
  (\overrightarrow{\psi_1})^2 = - \frac{t_1}{g}
\end{equation}
which exists and is stable for $t_1<0$.\\
The sublattice magnetizations:
\begin{eqnarray}\label{Eq19}
\overrightarrow{m}_1 &=&\frac{S_0}{\lambda_1} \overrightarrow{\psi_1}\\
\nonumber \overrightarrow{m}_2 &=&\frac{S_1}{\lambda_1} \overrightarrow{\psi_1}
\end{eqnarray}
will be antiparallel as  $S_1\sim K/D$ and $K<0$. The phase described by the above equations  will be presented by two antiparallel sublattices with different magnitudes of sublattice magnetizations.\\
In the weak coupling limit for antiparallel configuration, \emph{i.e.}, when $J_1>|K|, J_2>|K|$, or equivalently $J_1J_2 > K^2$, the system of equations~(\ref{Eq15}), together with the stability conditions (\ref{Eq16}), (\ref{Eq17}) should be solved. This can be done numerically and the results will be presented in the next section.

\section{\label{4} Results and discussion}

The analytical result for sublattice magnetizations in the limiting case of strong coupling~(\ref{Eq19}) gives for the magnitude of total magnetization $|\overrightarrow{M}| = |\overrightarrow{m}_1+ \overrightarrow{m}_2|$ the following expression:
$$|\overrightarrow{M}|=S_0 \frac{|\psi_1|}{\lambda_1}\left|1+\frac{S_1}{S_0}\right|$$ with $|\psi_1|^2$, given by (\ref{Eq18}). The phase transition is obviously of second order and the total magnetization behaviour with temperature is smooth resembling the one of Weiss ferromagnet with the exception that no saturation is reached for $T=0$.  According to the Neel's classification of ferrimagnets, see~\cite{Neel:1971}, the change of magnetization with temperature in the strong coupling limit falls within R-type curve. For example, similar curve is obtained theoretically and compared with the respective experiment for Y$_3$Fe$_5$O$_{12}$~\cite{Ranke:2009} where two sublattice model with strong antiferromagnetic coupling is considered.\\
In the limiting case when $J_1=J_2=J$ the relation will hold $S_0=-S_1= 1/\sqrt{2}$ and an antiferromagnetic structure with $\overrightarrow{m}_1= - \overrightarrow{m}_2$ will appear, only if $\overrightarrow{\psi}_2\equiv 0$ and $|\overrightarrow{\psi}_1|^2=-2t_1/u$. The transition temperature for antiferromagnetic ordering will be given by: $t_c^{a}= (J+|K|)/n$.\\
Further we  will present the numerical results for the temperature dependence of sublattice magnetizations and the total magnetization of the system in the  weak-coupling limit which we define here in the following way: $\mathcal{J}_1 > |\mathcal{K}|$ and $\mathcal{J}_2 > |\mathcal{K}|$.
Such a situation is present, for example in some ferrimagnetic compounds
like GdCo$_{12}$B$_6$~\cite{Isnard:2012}. It is experimentally found
that the exchange constants within sublattices are ferromagnetic and
larger than the antiferromagnetic coupling between the sublattices; moreover there the magnetic anisotropy is small.\\
Experiments for some R-T compounds where R is a rare earth element
and T is a transition element, show that the exchange in the transition
metal sublattice is leading in magnitude, while the exchange in the
rare-earth ion sublattice can be safely ignored and considered as
negligible. The
intersublattice interaction is also small see, for example,
DyFe$_5$Al$_7$~\cite{Gorbunov:2012},
ErFe$_{11}$TiH~\cite{Kostyuchenko:2015}, RCo$_2$ (R = Tb and Gd and R
= Er, Ho, and Dy)~\cite{Valiev:2017}. In our notations the relation
between the exchange integrals in this case will be
$\mathcal{J}_1>|\mathcal{K}|\gg \mathcal{J}_2$, so this does not
fall into our assumption of weak coupling and will not be considered
here.\\

In order to solve numerically the equations of state~(\ref{Eq15}) for weak coupling between sublattices we introduce the following
dimensionless parameters.
\begin{subequations}\label{Eq22}
\begin{align}
t&= \frac{T}{\mathcal{J}_1 + \mathcal{J}_2}, \\
\alpha&= \frac{\mathcal{J}_1 - \mathcal{J}_2}{\mathcal{J}_1 +
	\mathcal{J}_2},\\
\beta&= \frac{\mathcal{K}}{\mathcal{J}_1 + \mathcal{J}_2},
\end{align}
\end{subequations}
with $t$ -- the dimensionless temperature.
In the above expression we have supposed that $\mathcal{J}_1
>\mathcal{J}_2$ which in view of symmetry in interchanging the
sublattices does not limit the consideration; then $\alpha >0$ and
$\beta<0$ as $\mathcal{K}<0$.

The weak coupling between the sublattices, namely $ \mathcal{J}_1 >
|\mathcal{K}|$ and $\mathcal{J}_1 > |\mathcal{K}|$ may be expressed by
the parameters from~(\ref{Eq22}) by the relation:
$$
\alpha^2 + \beta^2 <1.
$$
The parameter $\alpha$ is a measure for the difference in exchange parameters of the two sublattices and by its definition $0<\alpha<1$.\\
The total magnetization of the system is the sum of sublattice
magnetizations~(\ref{Eq10}):
\begin{equation}\label{Eq23}
\overrightarrow{M}=\overrightarrow{m}_1+\overrightarrow{m}_2
=\frac{S_0+S_1}{\lambda_1}\overrightarrow{\psi}_1+\frac{S_0-S_1}{\lambda_2}\overrightarrow{\psi}_2.
\end{equation}
Hereafter we will use the following notations for magnitudes of sublattice magnetizations and total magnetization both in the text and in figures:
$$m_1=|\overrightarrow{m}_1|; \; \;m_2=|\overrightarrow{m}_2|; \mathrm{and} \;M=|\overrightarrow{M}|$$
The calculations show that the phase transition to ordered ferrimagnetic state occurs at temperature :
$$t_c=\frac{1}{6}(1+\sqrt{\alpha^2+\beta^2})$$
which grows when either the difference between the exchange interactions in sublattices grows, or when the antiferromagnetic coupling is bigger, or both. The phase transition from disordered to ordered phase  is of second order.\\
We want to note that within the exchange approximation used here for the regime of weak coupling defined above with the decrease of temperature a compensation point appears no matter how small is the difference between the  exchange interactions of sublattices. At the compensation temperature $t_{comp}$, the sublattice magnetizations $\overrightarrow{m}_1$ and $\overrightarrow{m}_2$ are equal in magnitude and antiparallel, so $M=0$. The relation between the order parameters magnitudes there is defined by:
\begin{equation}\label{Eq24}
\psi_1= \frac{\lambda_1}{\lambda_2}\frac{(S_0-S_1)}{(S_0+S_1)}\psi_2.
\end{equation}
As the calculations show $\psi_2< \psi_1$ for all values of $\alpha$
and $\beta$, but $\psi_1$ grows with the decrease of temperature in a
monotonic way, while $\psi_2$ grows more rapidly. The quantity
$$
\frac{\lambda_1}{\lambda_2}\frac{(S_0-S_1)}{(S_0+S_1)}=\frac{(6t_c)^2}{1-\alpha^2-\beta^2}\left(\frac{\sqrt{\alpha^2+\beta^2}-\beta}{\alpha}\right)
$$
is always $> 1$ as $\beta<0 \sim K$ and $\alpha >0$ so at some
temperature $t_\mathrm{comp}<t_c$, and values of $\psi_1 \;, \; \psi_2$ the
condition~(\ref{Eq24}) is fulfilled.
In the following figure we show the change of net magnetization magnitude $M (t)$ with the temperature for $\alpha = 0.1$, \textsl{i.e.}, $\mathcal{J}_2=0.83 \mathcal{J}_1$ and different values of $\beta$.
\begin{figure}[!ht]
\begin{center}
\includegraphics[scale=0.55]{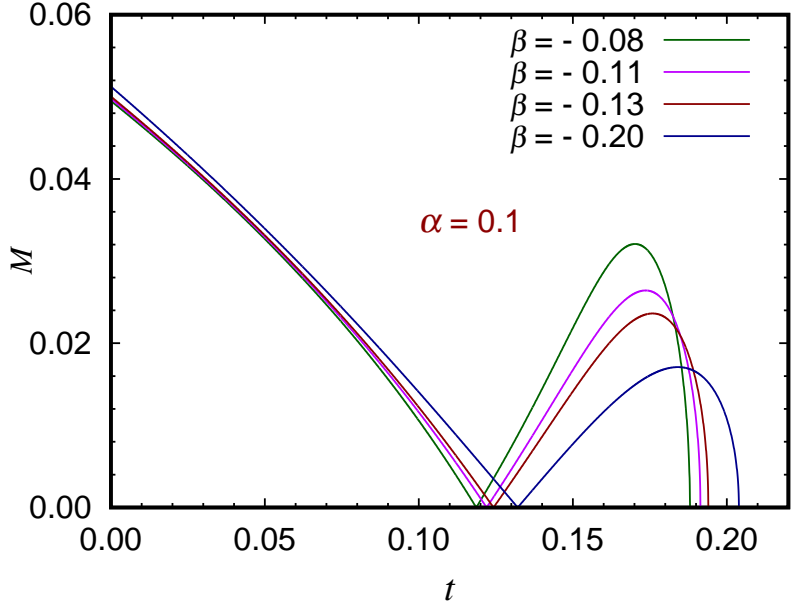}
\end{center}
\caption{The dependence of net magnetization $M$  on reduced
temperature $t$ for for fixed $\alpha$ and the different values of antiferromagnetic coupling $\beta$}. \label{Fig1}
\end{figure}
It is seen from Fig.1 that the increase of antiferromagnetic coupling between the sublattices slightly shifts the compensation temperature to higher values, and $M(t)$ grows more rapidly below the compensation temperature and reaches higher values as $t\longrightarrow 0$. We suppose that within the exchange approximation and in weak coupling limit the key factor for the compensation point to appear is the weakness of antiferromagnetic exchange between the sublattices compared to the ferromagnetic exchange  of sublattices 1 and 2 , respectively.

We will discuss in more detail the influence of difference between the magnitudes of exchange interaction in the sublattices, represented by the parameter $\alpha$ on $M(t)$ and sublattice magnetizations $m_1, \; \mathrm{and} \;m_2$.
For $\alpha=0.08$, \textsl{i.e.}, $\mathcal{J}_2 =0.85 \mathcal{J}_1$,  $M(t)$, $m_1, \; \mathrm{and} \;m_2$ are shown in Fig. 2.
 \begin{figure}[!ht]
\begin{center}
\includegraphics[scale=0.55]{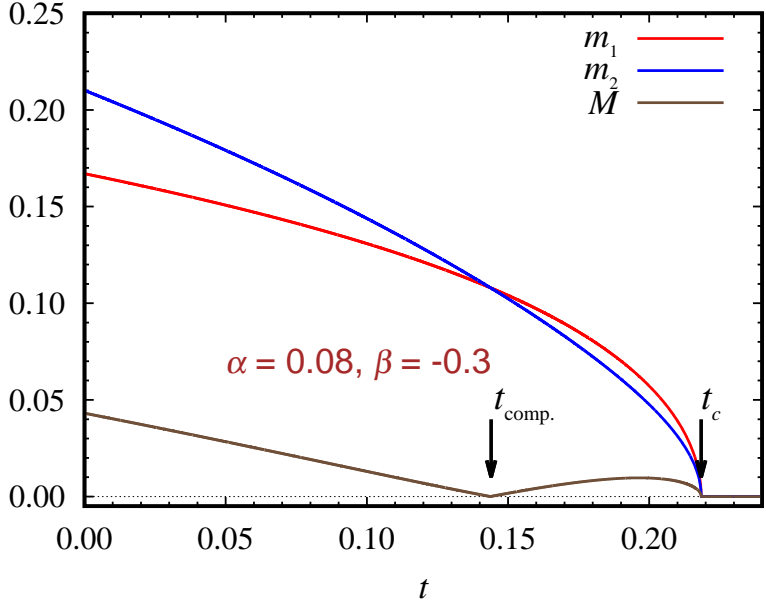}
\end{center}
\caption{The dependence of net magnetization $M$ and sublattice magnetizations $m_1$ and $m_2$ on reduced
temperature $t$ for for small difference between sublattice exchange parameters}. \label{Fig2}
\end{figure}
At $t_c$ the transition is of second order and when lowering the temperature  a compensation point $t_{comp}$ appears, which is located close to $t_c$.
Sublattice magnetizations change with temperature in  monotonic way, and in the temperature interval $t_{comp}<t<t_c$, the relation between  sublattice magnetizations is $m_1>m_2$, as expected as in sublattice 1 the  exchange interaction $\mathcal{J}_1>\mathcal{J}_2$. Below $t_{comp}$ the magnetization of weaker sublattice $m_2$ becomes bigger than $m_1$.\\
\begin{figure}[!ht]
\begin{center}
\includegraphics[scale=0.55]{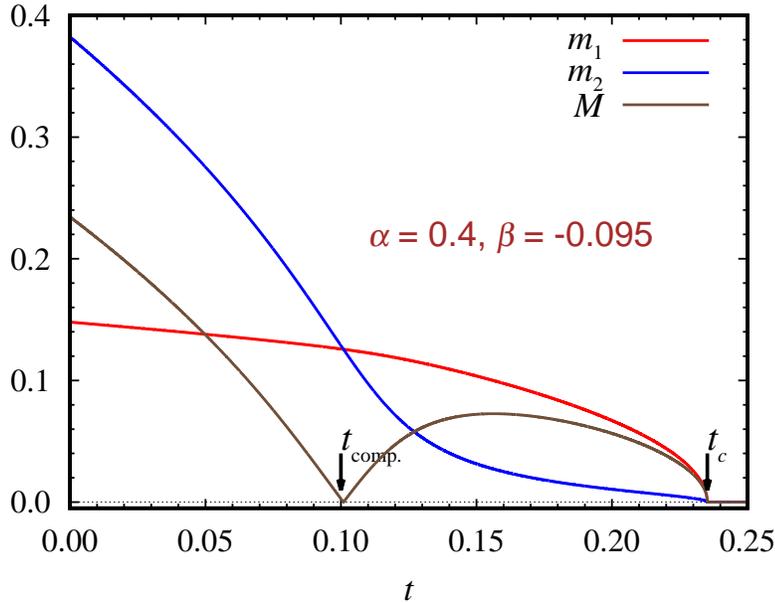}
\end{center}
\caption{The dependence of net magnetization $M$ and sublattice magnetizations $m_1$ and $m_2$ on reduced
temperature $t$ for intermediate difference between sublattice exchange parameters}. \label{Fig2}
\end{figure}
For intermediate values of $\alpha=0.4, \mathrm{or} \mathcal{J}_2 =0.43 \mathcal{J}_1$, see Fig.3,  with decrease of temperature  below the compensation point the total magnetization rapidly grows in non-monotonic way as $t\rightarrow 0$, similar to V-curve according to Neel's classification.
The behaviour of sublattice magnetizations with temperature is quite different; $m_1$  , which is the sublattice magnetization with stronger exchange interaction $\mathcal{J}_1$ grows in smooth way with decrease of temperature, while $m_2$ for weaker sublattice interaction $\mathcal{J}_2$, below compensation point grows drastically in non-monotonic way and in the temperature interval $0 <t<t_{comp}$, $m_2 \gg m_1$.\\
Such behaviour is described in detail in ~\cite{Belov:1996} for ferrimagnets with compensation point for many experimentally observed substances. There explanation of $M(t)$ behaviour below compensation point is done by introducing the notion of weak sublattice where depending on the particular substance considered different mechanisms for explanation of this effect are pointed out. Within our exchange model the effect of weak sublattice is readily seen and  mainly depends  on the difference $ \mathcal{J}_1-\mathcal{J}_2 \sim \alpha$. When $\alpha$ further grows the compensation point is shifted to lower temperatures and the magnetization $m_1$ of the stronger sublattice decreases for $t\rightarrow 0$. This is illustrated in Fig.4  for $ \mathcal{J}_2=0.19  \mathcal{J}_1$.\\
 \begin{figure}[!ht]
\begin{center}
\includegraphics[scale=0.55]{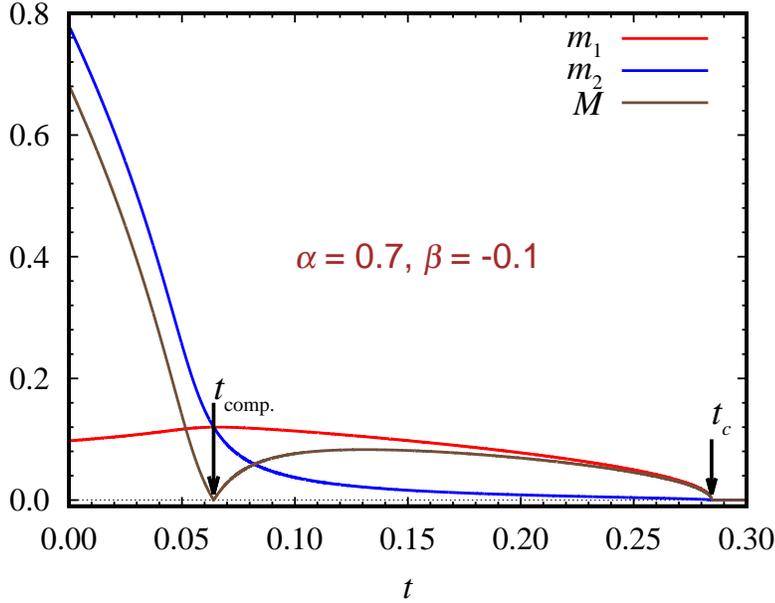}
\end{center}
\caption{Illustration of the dependence of net magnetization $M$ and sublattice magnetizations $m_1$ and $m_2$ on reduced
temperature $t$ when the exchange in sublattice 2 is very small compared to sublattice 1}. \label{Fig4}
\end{figure}

A qualitatively similar behaviour of the total magnetization can be seen in the
experiments with ErFe$_2$~\cite{Chaaba:2017} and
GdCo$_{12}$B$_6$~\cite{Isnard:2012} although direct comparison with
the experimental curves can hardly be made, as these substances have
crystallographic and magnetic structure that differs from the one
assumed within our model.

The influence of parameter $\alpha$ is summarized in the next figure, see Fig. 5, where the net magnetization is displayed for small values of $\beta=0.08$ and different values of parameter $\alpha$.
\begin{figure}[!ht]
\begin{center}
\includegraphics[scale=0.55]{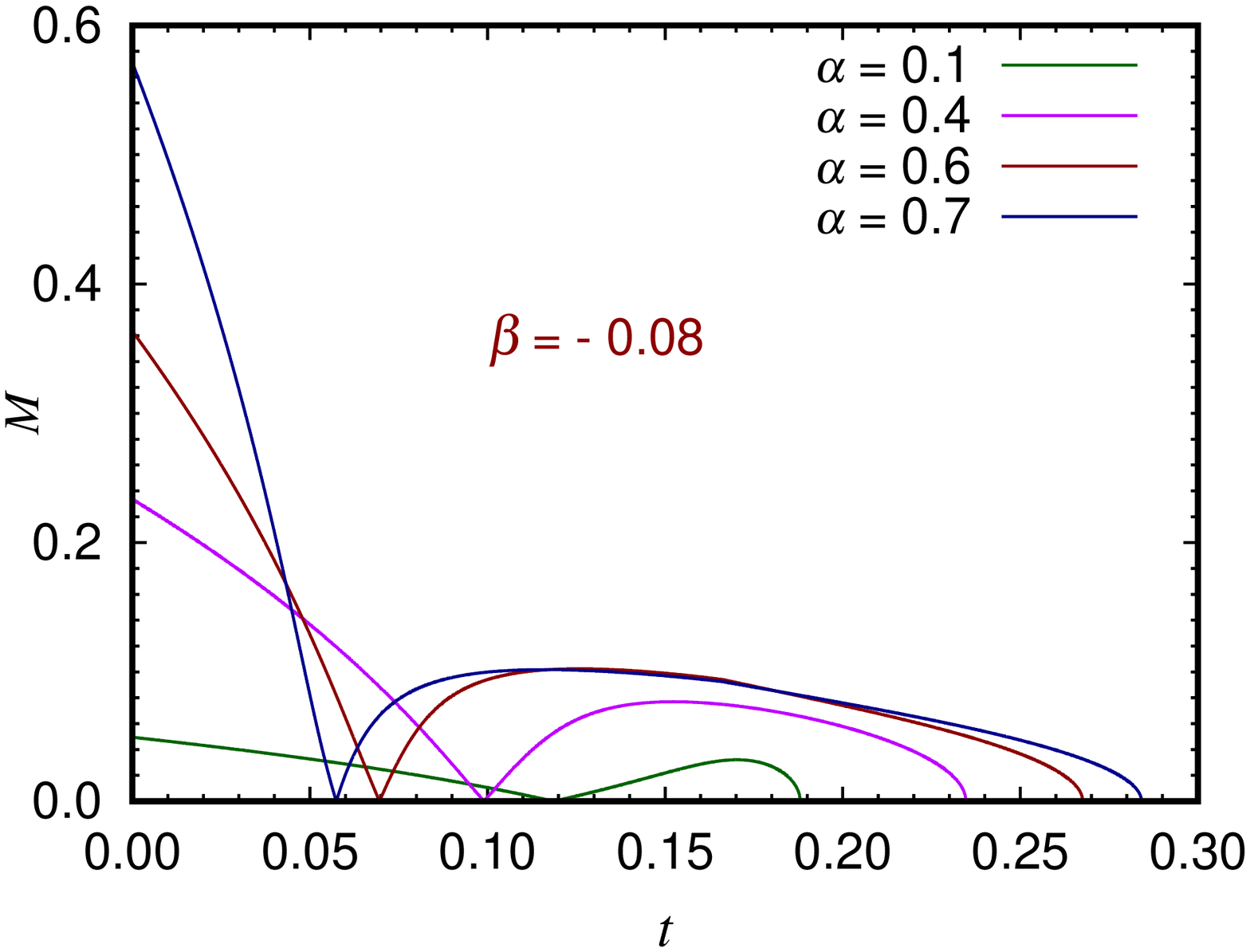}
\end{center}
\caption{The dependence of net magnetization $M$ for fixed antiferromagnetic exchange $\beta$ and growing difference $\alpha$  between the ferromagnetic exchanges of the two sublattices}. \label{Fig5}
\end{figure}
As the difference between the magnitudes of sublattice exchange interactions grow the transition temperature is shifted to higher values as expected and the compensation temperature is lowered.  Another effect is the change of net magnetization behaviour below $t_{comp}$ from monotonic to nearly exponential when $\alpha$ increases.

\section{\label{4} Conclusions}
Our mean-field analysis of this relatively simple model with  competing
interactions on a bcc lattice shows that the behaviour of net
magnetization of the two-sublattice ferrimagnet depends essentially on
the difference in magnetic interactions between the sublattices within
the weak coupling limit presented here. The influence of
antiferromagnetic coupling is more prominent when its magnitude is of
the same order, or larger than the difference between the exchange parameters of
the sublattices. The model may be generalized to include the expansion of
the effective Hamiltonian up to sixth-order terms in $x_i , \;\; i=1,2$ in
order to consider the possibility of first order phase transition to
ferrimagnetic state due to the influence of external parameters as pressure and change of concentration. \\Within our approach the parameters of Landau
energy can be directly related to the averaged microscopic exchange
interactions as described in section 2 , and this relation is quite simple in the case of high symmetry crystal structure as considered here.
It will be of interest also to include the influence of an external
magnetic field and to perform calculations for ferrimagnets, for which the
relation between exchange interactions fulfills the condition $\mathcal{J}_1 >
|\mathcal{K}| \gg \mathcal{J}_2$, that is, when one of the sublattices is very weak.

\section*{Acknowledgments}
This work was supported by the Bulgarian National Science Fund under
contract DN08/18 (14.12.2017).

\end{document}